\documentclass[letterpaper]{article}
\usepackage[utf8]{inputenc}
\usepackage{authblk}
\usepackage{graphicx}
\usepackage{a4wide}
\usepackage{subcaption}
\usepackage{amsmath,amssymb,amsthm,mathrsfs,amsfonts,xfrac,pifont} 

\usepackage{tabularx}
\usepackage{epigraph}
\AtBeginDocument{}

\usepackage[dvipsnames]{xcolor}
\usepackage{tikz}
\usepackage{hyperref}
\usepackage[backend=biber,style=numeric]{biblatex}  

\title{Modeling Alzheimer's Disease: From Memory Loss to Plaque \& Tangles Formation}
\author{Sai Nag Anurag Nangunoori\thanks{California High School, San Ramon, California }, Akshara Karthic Mahadevan\thanks{Independence High School, Frisco, Texas} \footnote{Joint First Author}}

\date{}

\begin{document}

\maketitle
\begin{abstract}
We employ the Hopfield model as a simplified framework to explore both the memory deficits and the biochemical processes characteristic of Alzheimer’s disease. By simulating neuronal death and synaptic degradation through increasing the number of stored patterns and introducing noise into the synaptic weights, we demonstrate hallmark symptoms of dementia, including memory loss, confusion, and delayed retrieval times. As the network’s capacity is exceeded, retrieval errors increase, mirroring the cognitive confusion observed in Alzheimer's patients. Additionally, we simulate the impact of synaptic degradation by varying the sparsity of the weight matrix, showing impaired memory recall and reduced retrieval success as noise levels increase. Furthermore, we extend our model to connect memory loss with biochemical processes linked to Alzheimer’s. By simulating the role of reduced insulin sensitivity over time, we show how it can trigger increased calcium influx into mitochondria, leading to misfolded proteins and the formation of amyloid plaques. These findings, modeled over time, suggest that both neuronal degradation and metabolic factors contribute to the progressive decline seen in Alzheimer's disease. Our work offers a computational framework for understanding the dual impact of synaptic and metabolic dysfunction in neurodegenerative diseases.
\end{abstract}
\section*{Introduction}
Alzheimer’s disease (AD) is a progressive neurodegenerative disorder characterized by the gradual decline in cognitive functions, including memory, learning, and reasoning. Despite extensive research, the mechanisms underlying Alzheimer’s disease remain incompletely understood. Two prominent features of the disease are the accumulation of extracellular amyloid plaques and intracellular neurofibrillary tangles, both of which are associated with widespread synaptic loss and neuronal death. These pathological changes lead to significant impairments in memory retrieval, cognitive confusion, and, ultimately, the loss of essential cognitive functions.

Theoretical models of memory and cognition, such as the Hopfield model, offer a powerful approach to exploring the dynamics of memory loss in Alzheimer's disease\cite{adeli2005alzheimer, herrmann1993neural, menschik1998neuromodulatory, cutsuridis2017computational, weber2017estimating}. The Hopfield model, originally designed as a simplified model for associative memory, has been widely used in neuroscience to study how networks of neurons store and retrieve information. By treating memory as an attractor state in a network of binary neurons, the Hopfield model allows us to simulate critical aspects of Alzheimer’s, including the degradation of memory storage and retrieval capabilities as synaptic connections deteriorate.

In this study, we extend the Hopfield model to explore not only memory degradation but also the role of metabolic factors in Alzheimer’s disease progression. Recent research \cite{DIEHL201726} suggests that metabolic dysfunctions, such as insulin resistance, are closely linked to neurodegenerative diseases, including Alzheimer's. Decreased insulin sensitivity, particularly in the brain, has been implicated in a cascade of harmful processes, including increased calcium influx into mitochondria, oxidative stress, and the formation of misfolded proteins that eventually accumulate as amyloid plaques. These plaques, in turn, contribute to further synaptic damage and neuronal loss.

Using a binary Hopfield network of densely connected neurons, we model both the cognitive and biochemical dimensions of Alzheimer's disease. We simulate memory loss by introducing noise to the synaptic weight matrix and demonstrate how this leads to impaired memory recall and slower retrieval times. Additionally, we investigate the effects of increasing neuronal load beyond the network’s theoretical capacity, showing how this can induce cognitive confusion, a symptom frequently observed in AD patients. Beyond cognitive symptoms, we incorporate biochemical simulations that link reduced insulin sensitivity with increased mitochondrial calcium influx, leading to protein misfolding and amyloid plaque formation.
\begin{figure}[h]
    \centering
    \includegraphics[width=0.6\textwidth]{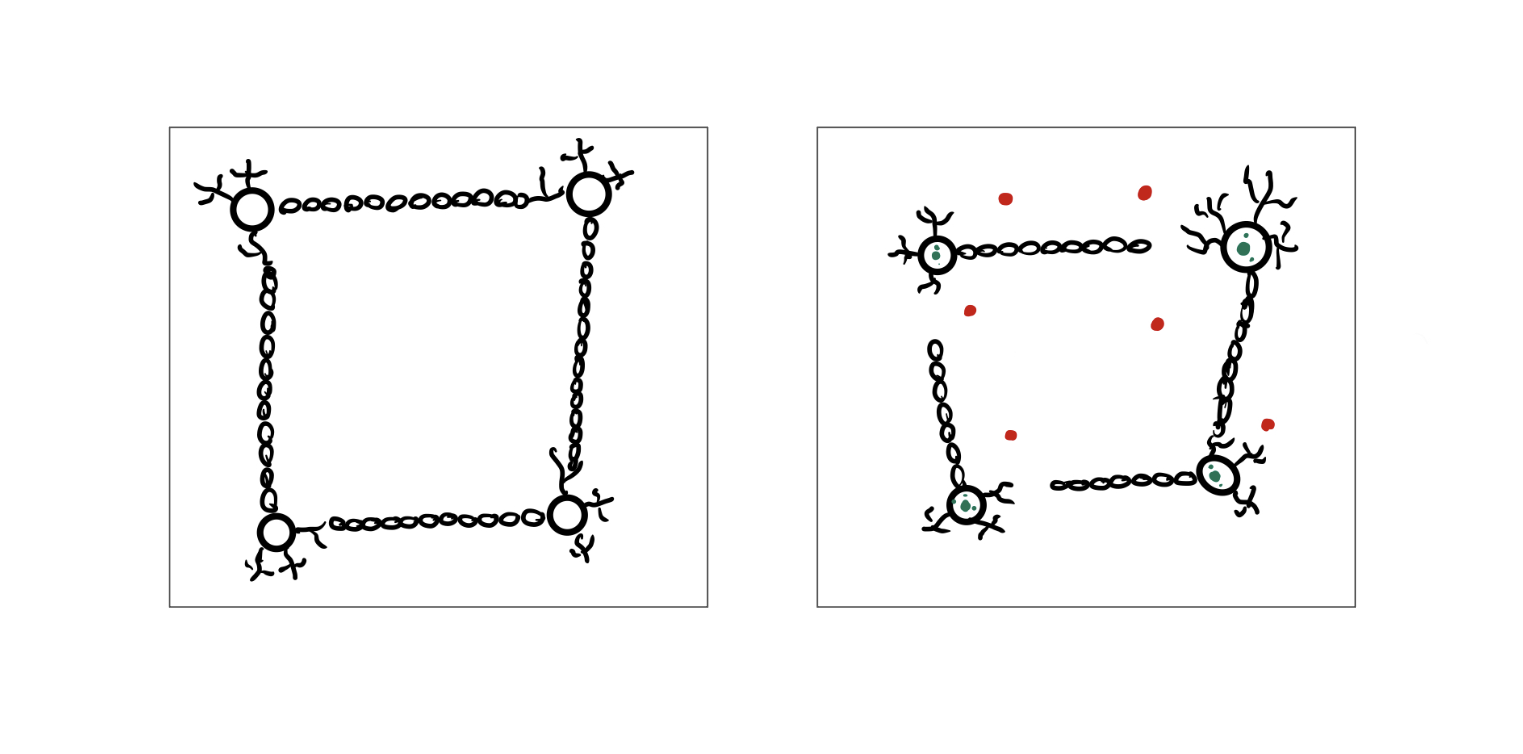}
    \caption{\textbf{Neuronal and Synaptic Degradation in Normal vs. Alzheimer’s Brain.} The left panel shows a representation of neurons in a normal brain, with intact synaptic connections. The right panel illustrates neurons in an Alzheimer’s brain, displaying hallmark pathological features, including red dots representing amyloid plaques and green dots representing neurofibrillary tangles. Synaptic loss is also highlighted, indicating the disruption in communication between neurons that is characteristic of Alzheimer’s disease.}
\end{figure}
This combined approach enables us to investigate Alzheimer’s disease from both a neurological and biochemical perspective, providing a comprehensive framework for understanding how synaptic degradation, neuronal death, and metabolic dysfunction contribute to the progressive decline in cognitive function. By integrating insights from associative memory models and metabolic processes, our work sheds light on the multifaceted mechanisms driving Alzheimer’s disease, with potential implications for the development of new therapeutic strategies.

\section*{Problem Setting}
Neural networks can store weakly correlated, associative patterns by following the prescription described by Hopfield in 1982\cite{hopfield1982neural}. A Hopfield neural network consists of N neurons which can store p patterns $\xi^\mu$ ($\mu=1,2,...,p$) that are $N\times1$ vectors with each element being either 1 or -1. A pattern is said to be stored if, when a weakly degraded version of it is presented to the network, say $I_{ext}$ which is $N\times1$ vector, then the network evolves to recovered form of that pattern as the final state. The learning of synaptic weights is via the Hebbian learning process according to which the change in synaptic weights is proportional to correlation in firing of pre- and post- synaptic neurons\cite{hebb1949organization}. 

After storing the patterns, the final weight matrix of the network is given by,
\begin{equation}
    J_{ij} = \frac{1}{N}\sum_{\mu=1}^{p}\xi_i^\mu\xi_j^\mu
\end{equation}
The neurons evolve synchronously by the following relation,
\begin{equation}
    S_i (t+1) = \rm{sgn}\left(\sum_{j=1}^{N}J_{ij}S_j(t)+I_{ext}\right)
\end{equation}
To quantify the total input neuron \textit{i} receives at time t, we introduce a local field variable, $h_{i}(t)$, that is defined as
\begin{equation}
    h_i (t) = \sum_{j=1}^{N}J_{ij}S_j(t)
\end{equation}
The network  evolves to a stationary state for all i when $h_{i}S_{i}>0$. The higher this value, the more stable the network is to noise. This fact will be used in the next section.
The original Hopfield model was able to accurately store and retrieve patterns only for full, symmetrically connected binary neurons storing uncorrelated patterns. Useful quantity of interest like maximum storage capacity was calculated in early works\cite{amit1985spin,amit1987statistical}. \par
The next few decades have been a long march towards making this original model more and more biologically plausible. The first step taken in this direction was to show robustness to dilution (deleting random connections) and noise in the network(\cite{derrida1987exactly,sompolinsky1986neural}). Then, the network was generalized to have sparse connectivity\cite{tsodyks1988enhanced}. In the 1990s, attractor dynamics in network of spiking neurons was analysed. \cite{amit1997dynamics,brunel2001effects}
Recently, the neurons have been generalized to be continuously varying rate units with sparse, asymmetric connectivity\cite{pereira2018attractor}. Lastly, it should be noted that the Hopfield prescription is not the only one for storing patterns in neural networks. There are models \cite{murray2017stable,mongillo2008synaptic} which provide alternative routes to short term memory maintenance. 

\section*{Results}
\subsection*{Impact of Synaptic Noise}
The Hopfield model provides a powerful framework for understanding associative memory and how it can degrade under certain conditions, such as those present in Alzheimer's disease. One of the key aspects of this model is its ability to store and retrieve patterns, where each pattern represents a particular memory or learned association. The capacity of the Hopfield network is limited by the number of neurons, with the theoretical maximum for reliable retrieval being approximately 0.15N, where N is the number of neurons. In our simulations, we used a network of 256 neurons, giving a theoretical capacity of approximately 35 stored patterns.

However, this capacity is highly sensitive to noise and synaptic degradation. Alzheimer's disease is characterized by the disruption of neuronal connections due to synaptic loss, plaques, and tangles, which can be modeled as noise introduced into the synaptic weights. The aim of the simulation presented in \textbf{Figure 2} is to investigate how increasing noise levels affect the network's ability to successfully retrieve stored patterns, reflecting memory degradation observed in Alzheimer’s patients.

In this simulation, we begin by storing a set of patterns in the network. To simulate memory retrieval under conditions of synaptic noise, we perturb the initial attractor states by adding noise scaled by different factors (c = 0.5, 1, 2, and 3). We then measure the retrieval success as a function of the number of stored patterns for each noise level. As shown in Figure 2, successful retrieval decreases more rapidly with higher noise levels, demonstrating that even moderate noise can significantly impair the network's performance. This decline is more pronounced as the number of stored patterns approaches and exceeds the network’s capacity, leading to a dramatic increase in retrieval failures, simulating the cognitive confusion associated with Alzheimer’s disease.

The results from this simulation provide important insights into how memory storage and retrieval processes break down as synaptic connections degrade, highlighting the critical role of synaptic integrity in preserving cognitive function. These findings suggest that noise introduced through synaptic degradation is a key contributor to the memory deficits seen in Alzheimer’s, as the brain struggles to correctly retrieve previously stored memories when faced with increasingly corrupted neural connections.
\begin{figure}[h]
    \centering
    \includegraphics[width=0.6\textwidth]{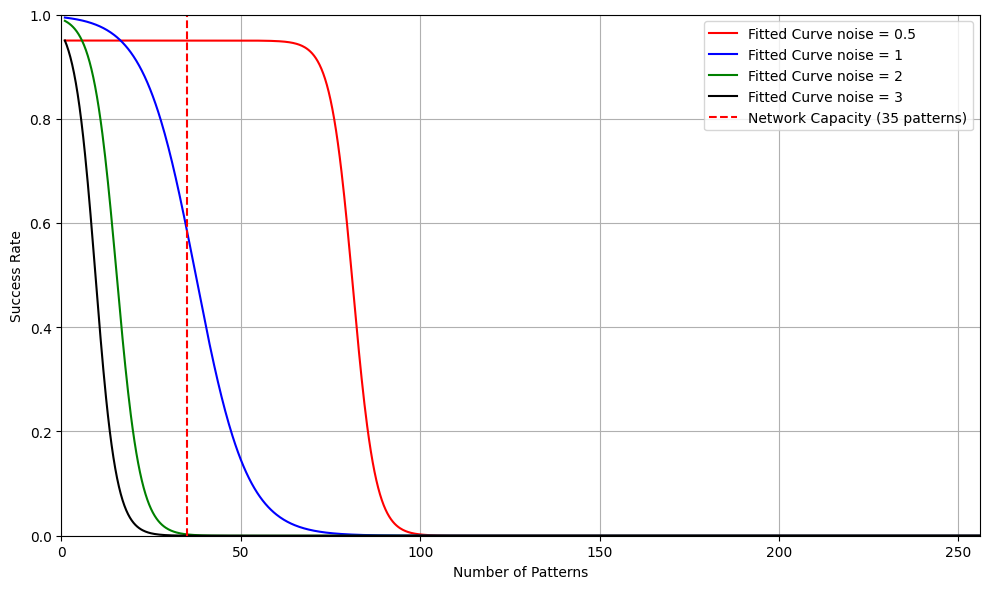}
    \caption{\textbf{Successful Memory Retrieval as a Function of Stored Patterns with Varying Noise Levels.}
The curves represent the best-fit retrieval probabilities for different levels of noise added to the initial attractor state, with noise levels c = 0.5, 1, 2, and 3. As the number of stored patterns increases, successful retrieval decreases more rapidly for higher noise levels. The theoretical network capacity, 35 patterns, is marked to indicate the maximum number of patterns that can be stored with high retrieval accuracy in a noise-free condition. Increased noise leads to earlier retrieval failures, simulating cognitive confusion as Alzheimer’s progresses.}
\end{figure}
\subsection*{Synaptic Sparsity and Its Effects on Memory Recall}
Synaptic degradation is a hallmark of Alzheimer’s disease and leads to the gradual breakdown of communication between neurons. In the context of memory, this synaptic loss impairs the brain's ability to retrieve stored information. In our model, synaptic degradation is simulated by introducing sparsity into the weight matrix of the Hopfield network, which mimics the loss of synaptic connections.

The simulation depicted in Figure 3 investigates the effects of synaptic sparsity on memory retrieval success. To model this degradation, we multiply the synaptic weight matrix by a sparsity matrix, where the probability f represents the fraction of intact connections (i.e., the probability of a connection being preserved between neurons). We simulate memory retrieval for different values of f, ranging from 0.9 (minimal degradation) to 0.5 (significant degradation), and measure the retrieval success as a function of the number of stored patterns.

As shown in \textbf{Figure 3}, memory retrieval success decreases as the value of f decreases. Higher synaptic sparsity, represented by lower values of f, leads to a more pronounced decline in retrieval accuracy as the number of stored patterns increases. This result aligns with the progressive memory loss observed in Alzheimer's patients, where synaptic degradation leads to an inability to retrieve memories correctly. Furthermore, the network’s capacity for storing patterns is sharply reduced as the synaptic connections become more sparse, leading to earlier retrieval failures as f decreases.

This simulation provides important insights into how synaptic loss, a key pathological feature of Alzheimer’s, directly impairs memory function. The results suggest that even moderate synaptic degradation can significantly affect memory retrieval efficiency, especially as the brain’s storage capacity is exceeded. By modeling synaptic loss in this way, we gain a clearer understanding of the mechanisms that lead to cognitive decline in neurodegenerative diseases, where synaptic sparsity disrupts the delicate balance required for effective memory storage and recall.

\begin{figure}[h]
    \centering
    \includegraphics[width=0.6\textwidth]{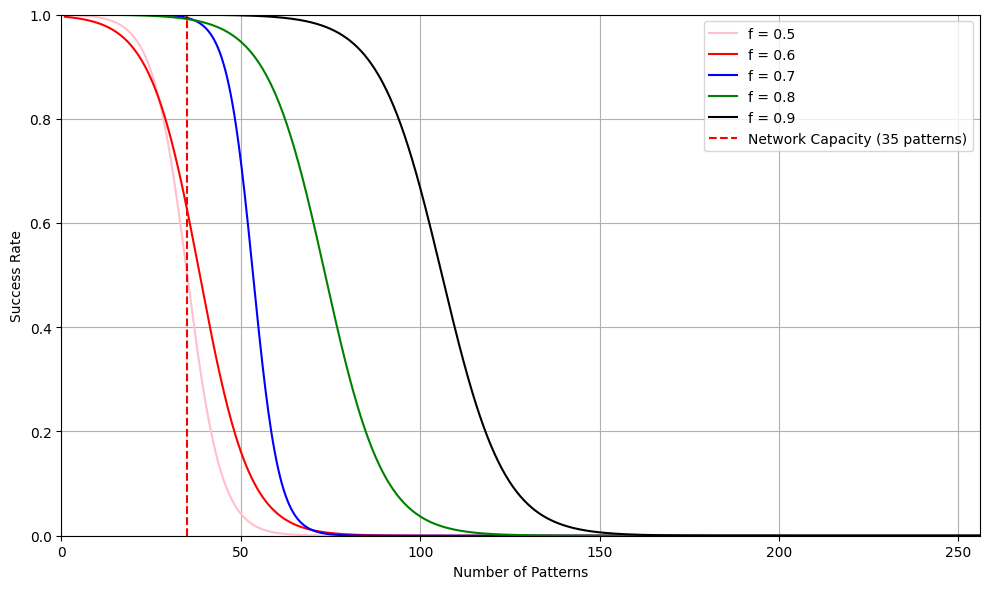}
    \caption{\textbf{Impact of Synaptic Degradation on Memory Retrieval as a Function of Stored Patterns.}
The degradation is simulated by multiplying the synaptic weight matrix by a sparsity matrix with probabilities f = 0.9, 0.8, 0.7, 0.6, and 0.5, where f represents the fraction of remaining synaptic connections. As the probability f decreases, indicating increased synaptic loss, the probability of successful memory retrieval declines more sharply with an increasing number of stored patterns. This result reflects how synaptic degradation in Alzheimer’s disease impairs memory function, leading to retrieval failure as neuronal connections weaken.}
    \label{fig:example}
\end{figure}
\subsection*{Modeling the Biochemical Pathways of Alzheimer’s: Insulin Sensitivity and Plaque Formation}
In addition to cognitive decline, Alzheimer’s disease is increasingly associated with metabolic dysfunction, particularly reduced insulin sensitivity. This metabolic impairment has been implicated in a cascade of biochemical events that exacerbate the progression of Alzheimer’s, including increased calcium influx into mitochondria, protein misfolding, and the eventual formation of amyloid plaques and neurofibrillary tangles.

The simulation shown in \textbf{Figure 4} explores how declining insulin sensitivity can lead to biochemical changes that contribute to Alzheimer’s pathology. Over time, as insulin sensitivity decreases, the influx of calcium from mitochondria-associated membranes (MAMs) into the mitochondria increases, creating conditions that foster oxidative stress and the misfolding of proteins. Misfolded proteins are precursors to the development of amyloid plaques, one of the primary pathological features of Alzheimer’s disease.

In this simulation, insulin sensitivity(IS) is modeled as decreasing over a decade, starting from normal levels and progressively declining. It is modelled as an inverted logistic function. The top left plot in \textbf{Figure 4} depicts this decrease in insulin sensitivity. The corresponding increase in mitochondrial calcium influx over time $Ca_{in}(t)$ is shown in the top right plot. It is given by the following equation:
\begin{equation*}
    Ca_{in}(t)=Ca_{0}+k_2\left(1-\frac{IS(t)}{IS_{max}}\right),
\end{equation*}
where $Ca_{0}$ is basal calcium influx and $k_2$ is the rate constant for increased calcium influx due to decreased insulin sensitivity. This increased calcium influx is directly linked to the formation of misfolded proteins, as demonstrated in the bottom left plot, which shows the accumulation of these proteins over time. The bottom right plot links these misfolded proteins M(t) to the formation of amyloid plaques P(t). It evolves according to the following equation,
\begin{equation*}
    \frac{dP}{dt}=k_7M(t)-k_8P(t),
\end{equation*}
where $k_7$ is the rate constant for plaque formation per unit misfolded protein and $k_8$ is  the rate constant for plaque clearance (often negligible in Alzheimer's disease). The values for all the parameters are given in the GitHub repository associated with the paper.
This simulation provides a mechanistic understanding of how metabolic factors, specifically insulin resistance, contribute to the biochemical processes underlying Alzheimer’s disease. By modeling the relationship between insulin sensitivity, calcium influx, and protein misfolding, we highlight the connection between metabolic dysfunction and neurodegeneration. This dual focus on both cognitive and metabolic pathways offers a more comprehensive framework for understanding the multifaceted nature of Alzheimer’s disease, with potential implications for the development of metabolic-based therapeutic interventions.
\begin{figure}[h]
    \centering
    \includegraphics[width=0.6\textwidth]{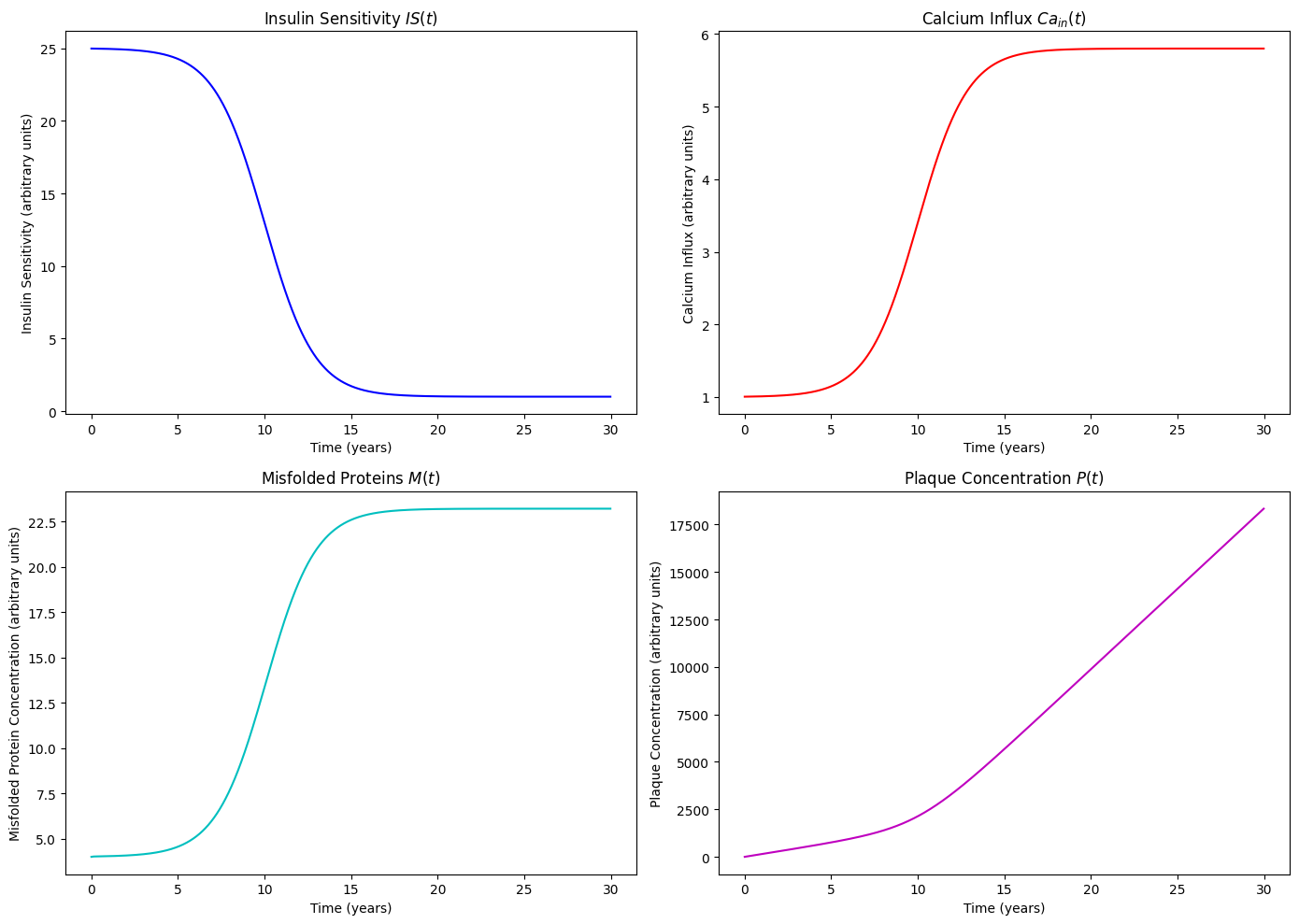}
    \caption{\textbf{Linking Insulin Sensitivity to Calcium Influx, Misfolded Proteins, and Plaque Formation Over Time.}
 The top left plot depicts the gradual decrease in insulin sensitivity over a decade, while the top right plot illustrates the corresponding increase in calcium influx into mitochondria from mitochondria-associated membranes (MAMs) as a function of time. The bottom left plot shows the accumulation of misfolded proteins over time, and the bottom right plot depicts the formation of amyloid plaques as a function of time. Together, these plots demonstrate how metabolic dysfunction, specifically reduced insulin sensitivity, can drive the biochemical processes leading to protein misfolding and plaque formation, contributing to the progression of Alzheimer’s disease.}
\end{figure}

\section*{Conclusion}
In this paper, we used a binary Hopfield model to simulate key aspects of Alzheimer’s disease, including both cognitive and biochemical features. Through the model, we demonstrated how memory loss, confusion, and slower retrieval times can arise as synaptic degradation and neuronal death occur. Our simulations showed that as the number of stored patterns increases and synaptic weights are increasingly disrupted by noise, memory retrieval becomes progressively less reliable. These findings align with the cognitive symptoms observed in Alzheimer’s patients, such as confusion and memory deficits as neuronal and synaptic function deteriorates.

Additionally, we extended our model to incorporate the role of metabolic dysfunction, specifically focusing on reduced insulin sensitivity, a factor that has been increasingly linked to Alzheimer's disease. We showed how decreased insulin sensitivity over time can lead to increased calcium influx into mitochondria, promoting the formation of misfolded proteins and amyloid plaques. This biochemical pathway mirrors the plaque and tangle formation seen in Alzheimer’s, linking metabolic and cognitive processes in the disease's progression.

Overall, our combined approach provides a simplified yet informative framework for understanding the multifaceted mechanisms underlying Alzheimer’s disease. By integrating synaptic degradation with metabolic dysfunction, our work offers new insights into the dual processes that drive both cognitive decline and biochemical changes in Alzheimer’s. These results highlight the importance of considering both neuronal and systemic factors when developing therapeutic strategies for neurodegenerative diseases like Alzheimer’s. Future research could expand on this model by incorporating more detailed biochemical dynamics or by exploring potential interventions that might mitigate both cognitive and metabolic decline.
\newline
\textbf{Code:} \href{https://github.com/Anurag-Nangunoori/Alzheimers_model}{github.com}

\end{document}